\pdfoutput=1   
\documentclass[aps,prl,reprint,superscriptaddress,showpacs,amsfonts,amssymb,amsmath]{revtex4-1} 
 
\usepackage{dcolumn} 
\usepackage{graphicx} 
\usepackage[colorlinks,breaklinks,linkcolor=red,anchorcolor=blue,citecolor=green]{hyperref} 
\usepackage{slashed} 
\usepackage{xspace} 
 
\newcommand{\MET}{\mbox{$\not \!\! E_T$}\xspace} 
\newcommand{\METVEC}{\mbox{$\not \!\! {\vec E}_T$}\xspace} 
\newcommand{\invfb}{\mbox{fb$^{-1}$}\xspace} 
\newcommand{\mtw}{\mbox{$M_T^W$}\xspace} 
\newcommand{\ttbar}{\mbox{$t\bar t$}\xspace} 
\newcommand{\ppbar}{\mbox{$p\bar p$}\xspace} 
\newcommand{\roots}{\mbox{$\sqrt{s}$}\xspace} 
\newcommand{\Vtb}{\mbox{$|V_\mathit{tb}|$}\xspace} 
\newcommand{\Vts}{\mbox{$|V_\mathit{ts}|$}\xspace} 
\newcommand{\Vtd}{\mbox{$|V_\mathit{td}|$}\xspace} 
\newcommand{\Wt}{\ensuremath{\mathit{\!Wt}}\xspace} 
\newcommand{\gev}{\mbox{GeV}\xspace} 
\newcommand{\gevc}{\mbox{GeV/$c$}\xspace} 
\newcommand{\gevcc}{\mbox{GeV/$c^2$}\xspace} 
\newcommand{\secvtx}{{\sc secvtx}\xspace} 
\newcommand{\powheg}{{\sc powheg}\xspace} 
\newcommand{\pythia}{{\sc pythia}\xspace} 
\newcommand{\alpgen}{{\sc alpgen}\xspace} 
 
\begin{document} 
 
\title{\boldmath Measurement of the Single Top Quark Production 
Cross Section and \Vtb \\ in Events with One Charged Lepton, Large 
Missing Transverse Energy, \\ and Jets at CDF \\} 
 
\affiliation{Institute of Physics, Academia Sinica, Taipei, Taiwan 11529, Republic of China}
\affiliation{Argonne National Laboratory, Argonne, Illinois 60439, USA}
\affiliation{University of Athens, 157 71 Athens, Greece}
\affiliation{Institut de Fisica d'Altes Energies, ICREA, Universitat Autonoma de Barcelona, E-08193 Bellaterra (Barcelona), Spain}
\affiliation{Baylor University, Waco, Texas 76798, USA}
\affiliation{Istituto Nazionale di Fisica Nucleare Bologna, \ensuremath{^{jj}}University of Bologna, I-40127 Bologna, Italy}
\affiliation{University of California, Davis, Davis, California 95616, USA}
\affiliation{University of California, Los Angeles, Los Angeles, California 90024, USA}
\affiliation{Instituto de Fisica de Cantabria, CSIC---University of Cantabria, 39005 Santander, Spain}
\affiliation{Carnegie Mellon University, Pittsburgh, Pennsylvania 15213, USA}
\affiliation{Enrico Fermi Institute, University of Chicago, Chicago, Illinois 60637, USA}
\affiliation{Comenius University, 842 48 Bratislava, Slovakia; Institute of Experimental Physics, 040 01 Kosice, Slovakia}
\affiliation{Joint Institute for Nuclear Research, RU-141980 Dubna, Russia}
\affiliation{Duke University, Durham, North Carolina 27708, USA}
\affiliation{Fermi National Accelerator Laboratory, Batavia, Illinois 60510, USA}
\affiliation{University of Florida, Gainesville, Florida 32611, USA}
\affiliation{Laboratori Nazionali di Frascati, Istituto Nazionale di Fisica Nucleare, I-00044 Frascati, Italy}
\affiliation{University of Geneva, CH-1211 Geneva 4, Switzerland}
\affiliation{Glasgow University, Glasgow G12 8QQ, United Kingdom}
\affiliation{Harvard University, Cambridge, Massachusetts 02138, USA}
\affiliation{Division of High Energy Physics, Department of Physics, University of Helsinki, FIN-00014 Helsinki, Finland; Helsinki Institute of Physics, FIN-00014 Helsinki, Finland}
\affiliation{University of Illinois, Urbana, Illinois 61801, USA}
\affiliation{The Johns Hopkins University, Baltimore, Maryland 21218, USA}
\affiliation{Institut f\"{u}r Experimentelle Kernphysik, Karlsruhe Institute of Technology, D-76131 Karlsruhe, Germany}
\affiliation{Center for High Energy Physics, Kyungpook National University, Daegu 702-701, Korea; Seoul National University, Seoul 151-742, Korea; Sungkyunkwan University, Suwon 440-746, Korea; Korea Institute of Science and Technology Information, Daejeon 305-806, Korea; Chonnam National University, Gwangju 500-757, Korea; Chonbuk National University, Jeonju 561-756, Korea; Ewha Womans University, Seoul 120-750, Korea}
\affiliation{Ernest Orlando Lawrence Berkeley National Laboratory, Berkeley, California 94720, USA}
\affiliation{University of Liverpool, Liverpool L69 7ZE, United Kingdom}
\affiliation{University College London, London WC1E 6BT, United Kingdom}
\affiliation{Centro de Investigaciones Energeticas Medioambientales y Tecnologicas, E-28040 Madrid, Spain}
\affiliation{Massachusetts Institute of Technology, Cambridge, Massachusetts 02139, USA}
\affiliation{University of Michigan, Ann Arbor, Michigan 48109, USA}
\affiliation{Michigan State University, East Lansing, Michigan 48824, USA}
\affiliation{Institution for Theoretical and Experimental Physics, ITEP, Moscow 117259, Russia}
\affiliation{University of New Mexico, Albuquerque, New Mexico 87131, USA}
\affiliation{The Ohio State University, Columbus, Ohio 43210, USA}
\affiliation{Okayama University, Okayama 700-8530, Japan}
\affiliation{Osaka City University, Osaka 558-8585, Japan}
\affiliation{University of Oxford, Oxford OX1 3RH, United Kingdom}
\affiliation{Istituto Nazionale di Fisica Nucleare, Sezione di Padova, \ensuremath{^{kk}}University of Padova, I-35131 Padova, Italy}
\affiliation{University of Pennsylvania, Philadelphia, Pennsylvania 19104, USA}
\affiliation{Istituto Nazionale di Fisica Nucleare Pisa, \ensuremath{^{ll}}University of Pisa, \ensuremath{^{mm}}University of Siena, \ensuremath{^{nn}}Scuola Normale Superiore, I-56127 Pisa, Italy, \ensuremath{^{oo}}INFN Pavia, I-27100 Pavia, Italy, \ensuremath{^{pp}}University of Pavia, I-27100 Pavia, Italy}
\affiliation{University of Pittsburgh, Pittsburgh, Pennsylvania 15260, USA}
\affiliation{Purdue University, West Lafayette, Indiana 47907, USA}
\affiliation{University of Rochester, Rochester, New York 14627, USA}
\affiliation{The Rockefeller University, New York, New York 10065, USA}
\affiliation{Istituto Nazionale di Fisica Nucleare, Sezione di Roma 1, \ensuremath{^{qq}}Sapienza Universit\`{a} di Roma, I-00185 Roma, Italy}
\affiliation{Mitchell Institute for Fundamental Physics and Astronomy, Texas A\&M University, College Station, Texas 77843, USA}
\affiliation{Istituto Nazionale di Fisica Nucleare Trieste, \ensuremath{^{rr}}Gruppo Collegato di Udine, \ensuremath{^{ss}}University of Udine, I-33100 Udine, Italy, \ensuremath{^{tt}}University of Trieste, I-34127 Trieste, Italy}
\affiliation{University of Tsukuba, Tsukuba, Ibaraki 305, Japan}
\affiliation{Tufts University, Medford, Massachusetts 02155, USA}
\affiliation{University of Virginia, Charlottesville, Virginia 22906, USA}
\affiliation{Waseda University, Tokyo 169, Japan}
\affiliation{Wayne State University, Detroit, Michigan 48201, USA}
\affiliation{University of Wisconsin, Madison, Wisconsin 53706, USA}
\affiliation{Yale University, New Haven, Connecticut 06520, USA}

\author{T.~Aaltonen}
\affiliation{Division of High Energy Physics, Department of Physics, University of Helsinki, FIN-00014 Helsinki, Finland; Helsinki Institute of Physics, FIN-00014 Helsinki, Finland}
\author{S.~Amerio\ensuremath{^{kk}}}
\affiliation{Istituto Nazionale di Fisica Nucleare, Sezione di Padova, \ensuremath{^{kk}}University of Padova, I-35131 Padova, Italy}
\author{D.~Amidei}
\affiliation{University of Michigan, Ann Arbor, Michigan 48109, USA}
\author{A.~Anastassov\ensuremath{^{v}}}
\affiliation{Fermi National Accelerator Laboratory, Batavia, Illinois 60510, USA}
\author{A.~Annovi}
\affiliation{Laboratori Nazionali di Frascati, Istituto Nazionale di Fisica Nucleare, I-00044 Frascati, Italy}
\author{J.~Antos}
\affiliation{Comenius University, 842 48 Bratislava, Slovakia; Institute of Experimental Physics, 040 01 Kosice, Slovakia}
\author{G.~Apollinari}
\affiliation{Fermi National Accelerator Laboratory, Batavia, Illinois 60510, USA}
\author{J.A.~Appel}
\affiliation{Fermi National Accelerator Laboratory, Batavia, Illinois 60510, USA}
\author{T.~Arisawa}
\affiliation{Waseda University, Tokyo 169, Japan}
\author{A.~Artikov}
\affiliation{Joint Institute for Nuclear Research, RU-141980 Dubna, Russia}
% ...

\collaboration{CDF Collaboration}
\altaffiliation[With visitors from]{
\ensuremath{^{a}}University of British Columbia, Vancouver, BC V6T 1Z1, Canada,
\ensuremath{^{b}}Istituto Nazionale di Fisica Nucleare, Sezione di Cagliari, 09042 Monserrato (Cagliari), Italy,
\ensuremath{^{c}}University of California Irvine, Irvine, CA 92697, USA,
\ensuremath{^{d}}Institute of Physics, Academy of Sciences of the Czech Republic, 182~21, Prague, Czech Republic,
\ensuremath{^{e}}CERN, CH-1211 Geneva, Switzerland,
\ensuremath{^{f}}Cornell University, Ithaca, NY 14853, USA,
\ensuremath{^{g}}University of Cyprus, Nicosia CY-1678, Cyprus,
\ensuremath{^{h}}Office of Science, U.S. Department of Energy, Washington, DC 20585, USA,
\ensuremath{^{i}}University College Dublin, Dublin 4, Ireland,
\ensuremath{^{j}}ETH, 8092 Z\"{u}rich, Switzerland,
\ensuremath{^{k}}University of Fukui, Fukui City, Fukui Prefecture, Japan 910-0017,
\ensuremath{^{l}}Universidad Iberoamericana, Lomas de Santa Fe, M\'{e}xico C.P. 01219, Distrito Federal, Mexico
\ensuremath{^{m}}University of Iowa, Iowa City, IA 52242, USA,
\ensuremath{^{n}}Kinki University, Higashi-Osaka City, Japan 577-8502,
\ensuremath{^{o}}Kansas State University, Manhattan, KS 66506, USA,
\ensuremath{^{p}}Brookhaven National Laboratory, Upton, NY 11973, USA,
\ensuremath{^{q}}Queen Mary, University of London, London, E1 4NS, United Kingdom,
\ensuremath{^{r}}University of Melbourne, Victoria 3010, Australia,
\ensuremath{^{s}}Muons, Inc., Batavia, IL 60510, USA,
\ensuremath{^{t}}Nagasaki Institute of Applied Science, Nagasaki 851-0193, Japan,
\ensuremath{^{u}}National Research Nuclear University, Moscow 115409, Russia,
\ensuremath{^{v}}Northwestern University, Evanston, IL 60208, USA,
\ensuremath{^{w}}University of Notre Dame, Notre Dame, IN 46556, USA,
\ensuremath{^{x}}Universidad de Oviedo, E-33007 Oviedo, Spain,
\ensuremath{^{y}}CNRS-IN2P3, Paris, F-75205 France,
\ensuremath{^{z}}Universidad Tecnica Federico Santa Maria, 110v Valparaiso, Chile,
\ensuremath{^{aa}}The University of Jordan, Amman 11942, Jordan,
\ensuremath{^{bb}}Universite catholique de Louvain, 1348 Louvain-La-Neuve, Belgium,
\ensuremath{^{cc}}Bergische Universit\"{a}t Wuppertal, 42097 Wuppertal, Germany,
\ensuremath{^{dd}}University of Z\"{u}rich, 8006 Z\"{u}rich, Switzerland,
\ensuremath{^{ee}}Massachusetts General Hospital, Boston, MA 02114 USA,
\ensuremath{^{ff}}Harvard Medical School, Boston, MA 02114 USA,
\ensuremath{^{gg}}Hampton University, Hampton, VA 23668, USA,
\ensuremath{^{hh}}Los Alamos National Laboratory, Los Alamos, NM 87544, USA,
\ensuremath{^{ii}}Universit\`{a} degli Studi di Napoli Federico I, I-80138 Napoli, Italy
}
\noaffiliation

\date{December 31, 2014} 
 
\begin{abstract}  
 
\noindent We report a measurement of single top quark production in
proton-antiproton collisions at a center-of-mass energy of \roots~=
1.96~TeV using a data set corresponding to 7.5~\invfb of integrated
luminosity collected by the Collider Detector at Fermilab. We select
events consistent with the single top quark decay process $t
\rightarrow Wb \rightarrow \ell\nu b$ by requiring the presence of an
electron or muon, a large imbalance of transverse momentum indicating
the presence of a neutrino, and two or three jets including at least
one originating from a bottom quark. An artificial neural network is
used to discriminate the signal from backgrounds. We measure a single
top quark production cross section of $3.04^{+0.57}_{-0.53}$~pb and
set a lower limit on the magnitude of the coupling between the top
quark and bottom quark $\Vtb > 0.78$ at the 95\% credibility level.
 
\end{abstract} 
 
 
\maketitle 
 
In the standard model (SM) of fundamental interactions, top quarks are
produced in hadron collisions primarily as top-antitop (\ttbar) pairs via
the strong interaction. The top quark was first observed in this production
mode in 1995~\cite{PhysRevLett.74.2626,*PhysRevLett.74.2632}. The top quark
is also produced singly via weak charged-current interactions. At the
Fermilab Tevatron proton-antiproton ($p\bar{p}$) collider, single top quark
production proceeds via the exchange of a virtual $W$ boson in the $t$
channel, via the decay of an intermediate $W$ boson in the $s$ channel, or
in association with a $W$ boson (\Wt)~\cite{ccmodes}. The respective SM
production cross sections at the Tevatron, calculated at approximate
next-to-next-to-leading-order accuracy in the strong coupling
$\alpha_s$, are $\sigma_{t} \approx 2.10$~pb~\cite{PhysRevD.83.091503},
$\sigma_{s} \approx1.06$~pb~\cite{PhysRevD.81.054028}, and $\sigma_{\Wt}
\approx 0.25$~pb~\cite{PhysRevD.82.054018} for a top quark mass of
172.5~\gevcc.
 
The measurement of the single top quark production cross section provides a
test of the SM via a direct determination of the magnitude of the
Cabibbo-Kobayashi-Maskawa (CKM)~\cite{PhysRevLett.10.531,*Kobayashi:1973fv}
matrix element $\Vtb$, as the cross section is proportional to $\Vtb^2$. The
strength of the coupling $\Vtb$ governs the decay rate of the top quark and
its decay width into $Wb$. As this measurement assumes only that $\Vtb^2 \gg
\Vts^2 + \Vtd^2$ and does not rely on an assumption about the unitarity of
the CKM matrix, it can constrain various extensions of the SM, namely models
with fourth-generation quarks, models with flavor-changing neutral currents,
and other phenomena not predicted by the SM~\cite{PhysRevD.63.014018}.
 
Single top quark production in the combined $s$ + $t$ channels was first
observed independently by the CDF and D0 experiments in
2009~\cite{PhysRevLett.103.092002,PhysRevLett.103.092001}. The D0
Collaboration updated its measurement in 2011~\cite{PhysRevD.84.112001} and
reported the observation of single top quark production in the $t$
channel~\cite{Abazov2011313}. The ATLAS and CMS experiments at the Large
Hadron Collider (LHC) reported measurements of single top quark production
in the $t$ channel in 2012~\cite{Chatrchyan:2012aa,*Aad2012330}. More
recently, these experiments presented evidence of single top quark
production via the $\Wt$ process~\cite{PhysRevLett.110.022003,*Aad2012142},
and CMS recently reported the observation of single top quark
production via this process~\cite{PhysRevLett.112.231802}. In addition, the
CDF and D0 experiments separately reported evidence for $s$-channel production
\cite{PhysRevLett.112.231804,*PhysRevLett.112.231805,*j.physletb.2013.09.048}
and jointly reported the observation of $s$-channel single top quark
production in 2014~\cite{PhysRevLett.112.231803}. The $s$-channel process is
difficult to observe at the LHC due to the small signal-to-background ratio.
 
In this Letter we report precise measurements of the single top quark
production cross section for (i) the sum of the $s$-channel, $t$-channel,
and \Wt processes, (ii) the $s$-channel process alone, and (iii) the sum of
the $t$-channel and \Wt processes, using more than twice the data of the
previous CDF measurement~\cite{PhysRevLett.103.092002,PhysRevD.82.112005}.
Using the measured single top quark cross section for the sum of
$s$-channel, $t$-channel, and \Wt processes, we also set a lower limit on
the coupling \Vtb. The data sample was collected at the Tevatron at a
center-of-mass energy of \roots~= 1.96~TeV. The data sample corresponds to
an integrated luminosity of 7.5~\invfb collected with the CDF II detector,
which includes a solenoidal magnetic spectrometer surrounded by
projective-geometry sampling calorimeters and muon
detectors~\cite{PhysRevD.71.032001}.

Since the magnitude of the top-bottom quark coupling is much larger than
that of the top-down and top-strange quark couplings, we assume that every
top quark decays into a $W$ boson and a bottom ($b$) quark. We identify
single top quark candidates by searching for the decay of a $W$ boson to a
neutrino and either an electron ($e$) or a muon ($\mu$). Candidate events are
required to have an $e$ or $\mu$ with large transverse momentum
$p_T$~\cite{coord}, a large imbalance in the event's total transverse
momentum (missing energy) \MET~\cite{cmet2} indicating a neutrino, and two
or three hadronic jets.
 
Events are collected by three sequential levels of online selection
requirements (triggers). We include events collected by high-$p_T$ lepton
triggers, where the candidate $e$ ($\mu$) has $E_T$~$>$ 20~\gev
($p_T$~$>$~18~\gevc) and pseudorapidity $|\eta| < 1.0$~\cite{coord}. We also utilize
novel triggers that require either \MET $>$ 35~GeV plus two jets or \MET $>$
45~GeV, which increase the acceptance by adding new types of identified muon
candidates~\cite{PhysRevD.82.112005}. Based on the type of lepton
identified, events are grouped into two mutually exclusive categories called
the tight lepton category and the extended muon category.

The final event selection requires a single isolated charged lepton with
$|\eta|<1.6$ and $p_T > 20~\gevc$, consistent with the leptonic decay of a
$W$ boson. After correcting \MET for the presence of jets and muons in the
event, we require $\MET > 25~\gev$ to reduce the background from multijet
events that do not contain a $W$ boson, referred to as the non-$W$
background. Jets are reconstructed using a fixed-cone
algorithm~\cite{Bhatti:2005ai} with radius $\Delta R = 0.4$ in
$\eta$-$\phi$ space~\cite{dr}. We select events with either two or three
jets having $E_T~> 20~\gev$ and $|\eta| <2.8$. In order to improve the
separation of signal from background, at least one of the jets must be
identified as originating from a $b$ quark (``$b$ tagged'') using the
{\secvtx} algorithm~\cite{PhysRevD.86.032011}.

Backgrounds that mimic the single top quark signal originate from events in
which a $W$ boson is produced in association with one or more heavy-flavor
jets ($W$ + \textit{HF}), events with light-flavor jets that are mistakenly
$b$ tagged ($W$ + \textit{LF}), multijet (non-$W$) events, \ttbar events,
diboson ($WW$, $W\!Z$, $Z\!Z$) events, and events with a $Z$ boson and jets.
In addition to the \MET requirement, we further reduce the non-$W$
background by using a dedicated selection that exploits the $W$ boson
transverse mass \mtw~\cite{mt} and the missing transverse energy
significance~\cite{PhysRevD.82.112005}. Events with a reconstructed muon in
the tight lepton category are required to have $\mtw > 10~\gevcc$, while the
remaining muon events and events triggered by an electron must have $\mtw >
20~\gevcc$.

Backgrounds are estimated using both data-driven algorithms and simulated
data from Monte Carlo (MC) samples~\cite{PhysRevD.82.112005}. The diboson
and $t\bar{t}$ processes are modeled using \pythia~\cite{PYTHIA}, and the
production of a $W$ or $Z$ boson associated with jets is modeled using
\alpgen~\cite{alpgen}. The single top quark signal is modeled using
\powheg~\cite{POWHEG2009,*Re:2010bp} at next-to-leading-order (NLO) accuracy
in $\alpha_s$. This measurement uses a NLO generator for the first time to
model $s$- and $t$-channel single top quark production with the proper
inclusion of the \Wt contribution~\cite{thesis}. A top quark mass of
172.5~GeV/$c^2$ is assumed, which is fully consistent with recent
measurements~\cite{PhysRevD.86.092003,*worldcombo}. Each of the event
generators uses the \textsc{cteq5l} parton distribution
functions~\cite{CTEQ5L} as input except for \powheg, which uses the
\textsc{cteq6.1} parton distribution functions~\cite{CTEQ61}. Parton
showering and hadronization is simulated using \pythia tuned to underlying
event data from the Tevatron~\cite{PhysRevD.82.034001}. The CDF~II detector
response is modeled using \textsc{geant3}~\cite{Geant3}.

The probability for a light-flavor jet to be mistakenly $b$ tagged is
estimated using a mistag matrix extracted from data control samples and
parametrized as a function of the jet and event
properties~\cite{PhysRevD.82.112005}. The kinematic properties of non-$W$
events are determined using data samples obtained with less stringent
requirements applied to lepton identification and isolation. The sample of
events prior to $b$ tagging, referred to as the pretag sample, is dominated
by non-$W$ and $W$ + jets events. As non-$W$ events typically have smaller
\MET than $W$ boson events, the normalization for both non-$W$ and $W$ +
jets events is determined by fitting the \MET distribution with the
\MET~$>$~25~GeV requirement removed.

Table~\ref{table:m2table} shows the expected sample composition for events
with either two or three jets and either one or two $b$ tags, corresponding
to a total of four statistically independent signal regions. Events
originating from $s$-channel single top quark production frequently populate
the two-tag signal region, while $t$-channel and $\Wt$ events predominantly
populate the one-tag signal region.
  
\newcolumntype{s}{D{,}{\mbox{ $\pm$ }}{-1}} 
\newcommand{\mc}{\multicolumn} 
\begin{table*}[htb!] 
\caption{Predicted and observed number of events in four statistically
independent signal regions, which consist of $W$ boson events with either
two or three jets, each with either one or two $b$ tags. The uncertainties
on the predictions include statistical and systematic contributions from
simulated samples and data-driven algorithms, as described in the
text.}\label{table:m2table}
\begin{center} 
\begin{tabular}{lssss} 
\hline \hline 
Process & \mc{1}{c}{\quad $W$ + 2 jets, 1 tag \quad} & \mc{1}{c}{\quad $W$ + 3 jets, 1 tag \quad} &  
\mc{1}{c}{\quad $W$ + 2 jets, 2 tags \quad} & \mc{1}{c}{\quad $W$ + 3 jets, 2 tags \quad} \\ 
\hline 
$t \bar t$ & 474,49 & 1067,109 & 98,14 & 284,42 \\ 
$WW$ & 148,21 & 48,7 & 1.1,0.3 & 1.2,0.3 \\ 
$W\!Z$ & 53,6 & 14,2 & 8.8,1.3 & 2.4,0.4 \\ 
$ZZ$ & 1.7,0.2 & 0.7,0.1 & 0.3,0.0 & 0.1,0.0 \\ 
$Z$ + jets & 118,15 & 46,6 & 4.8,0.7 & 2.7,0.4 \\ 
$W + b\bar{b}$ & 1452,437 & 434,131 & 183,56 & 65,20 \\ 
$W + c\bar{c}$ & 766,233 & 254,77 & 10 ,3 & 7,2 \\ 
$W + cj$ & 583,177 & 128,39 & 7.8,2.4 & 3.5,1.1 \\ 
$W$ + \textit{LF} & 1459,148 & 433,47 & 7.4,1.5 & 5.4,1.1 \\ 
non-$W$ & 316,126 & 141,57 & 6.8,3.5 & 3.4,3.2 \\ 
\hline 
$t$ channel & 193,25 & 84,11 & 6,1 & 15,2 \\ 
$s$ channel & 128,11 & 43,4 & 32,4 & 12 ,2 \\ 
$\Wt$  & 16,4 & 26,7 & 0.7,0.2 & 2.3,0.6 \\ 
\hline 
Total prediction & 5707,877 & 2719,293 & 367,66 & 403,53 \\ 
Observed & \mc{1}{c}{\quad$\,$ 5533} & \mc{1}{c}{\quad$\,$ 2432} & \mc{1}{c}{\quad$\:$ 335} & \mc{1}{c}{\quad$\:$ 355} \\ 
\hline \hline 
\end{tabular} 
\end{center} 
\end{table*} 
 
The number of expected signal events is much smaller than the uncertainty on
the predicted background, and further separation of signal and background is
required. We use artificial neural networks (NN)~\cite{Grosse:2004by} to
separate signal events from background events. Two dedicated NNs are used
for each of the four signal regions, one for each of the two lepton
categories. A number of kinematic variables are studied, and the most
significant for distinguishing signal from background are used as inputs to
build the NN discriminants. Although the NN inputs with the greatest
discriminating power vary for the four signal regions, examples of the best
inputs include $Q \times \eta$, the product of the charge of the electron or
muon and the pseudorapidity of the light-quark jet, and $M_{\ell\nu b}$, the
reconstructed top quark mass based on the electron or muon, the
reconstructed neutrino, and the $b$-tagged jet. Descriptions of the
variables and the full optimization procedure can be found in
Ref.~\cite{PhysRevD.82.112005}.
 
In order to maximize signal sensitivity, the NN discriminant is trained
using only $s$-channel events as signal in the two-jet, two-tag signal
region. In the remaining jet and tag signal regions, the NN is trained
assuming only $t$-channel events as signal. To further improve the precision
of the cross section measurement, we use training samples that contain
additional events in which the jet energy scale, renormalization scale, and
factorization scale are varied within their systematic uncertainties.  By training the NN 
with a broader set of events with features that more closely resemble data, 
the NN better accommodates certain systematic variations, enhancing its ability 
to discriminate signal from background. 
Simulations predict that this new procedure 
improves the accuracy of the final cross section measurement by 
approximately 3\%~\cite{thesis}.
 
\begin{figure}[tb]  
\begin{center}  
\includegraphics[width=1.0\linewidth]{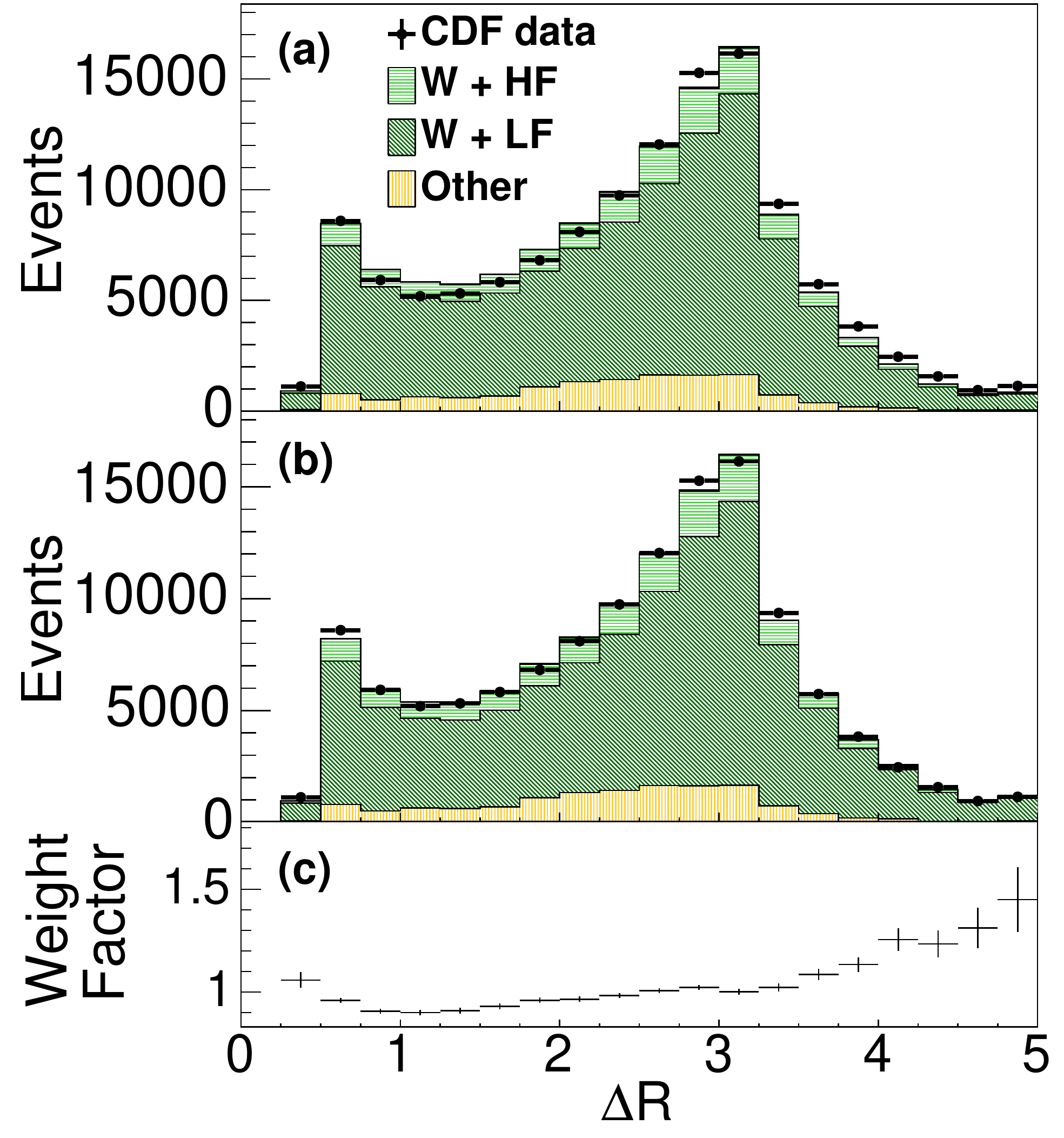}  
\label{fig:delr}  
\caption{\label{fig:deltaR} Distribution of $\Delta R(\vec{j_1},\vec{j_2})$ for the two 
most energetic jets in the $W$~+ 2~jet pretag sample (a) before reweighting and 
(b) after reweighting.  (c) The weight factor, which ranges up to 1.4 at large $\Delta R$.} 
\end{center}  
\end{figure}

The measurement of the single top quark cross section requires substantial
input from theoretical models, Monte Carlo simulations, and extrapolations
from control samples in data. We assign systematic uncertainties to the
predictions and we investigate the effects of these uncertainties on the
measured cross section. Three different classes of uncertainty are
considered: the uncertainty in the predicted rates of signal and background
processes, the uncertainty in the shapes of the distributions of the
discriminant variables, and the uncertainty arising from the simulated
sample size in each bin of each discriminant distribution. 
In the pretag sample, discrepancies between data and the MC predictions are 
visible in certain regions of distributions such as jet $\eta$ and $\Delta
R(\vec{j_1},\vec{j_2}) = \sqrt{(\Delta\eta)^2 + (\Delta\phi)^2}$, where
$\vec{j_1}$ and $\vec{j_2}$ are the momentum vectors of the two most
energetic jets~\cite{PhysRevD.82.112005} (see Fig.~\ref{fig:deltaR}). 
The inaccurate modeling of these
distributions is potentially significant because jet-related variables are important inputs to the 
NN. We determine that the mismodeling is mainly due to $W$~+~\textit{LF} 
events, and we account for the mismodeling by 
creating a modified $W$~+~\textit{LF} background template in which events 
are reweighted to match pretag data in the jet $E_{T}$, jet $\eta$, and 
$\Delta \phi(\vec{j_1},\vec{j_2})$ distributions~\cite{thesis}. The 
difference between the unweighted and weighted distributions 
is taken as a one-sided systematic uncertainty. All of the systematic 
uncertainties are thoroughly discussed in
Ref.~\cite{PhysRevD.82.112005}.

\begin{figure}[tb]  
\begin{center}  
\includegraphics[width=1.0\linewidth]{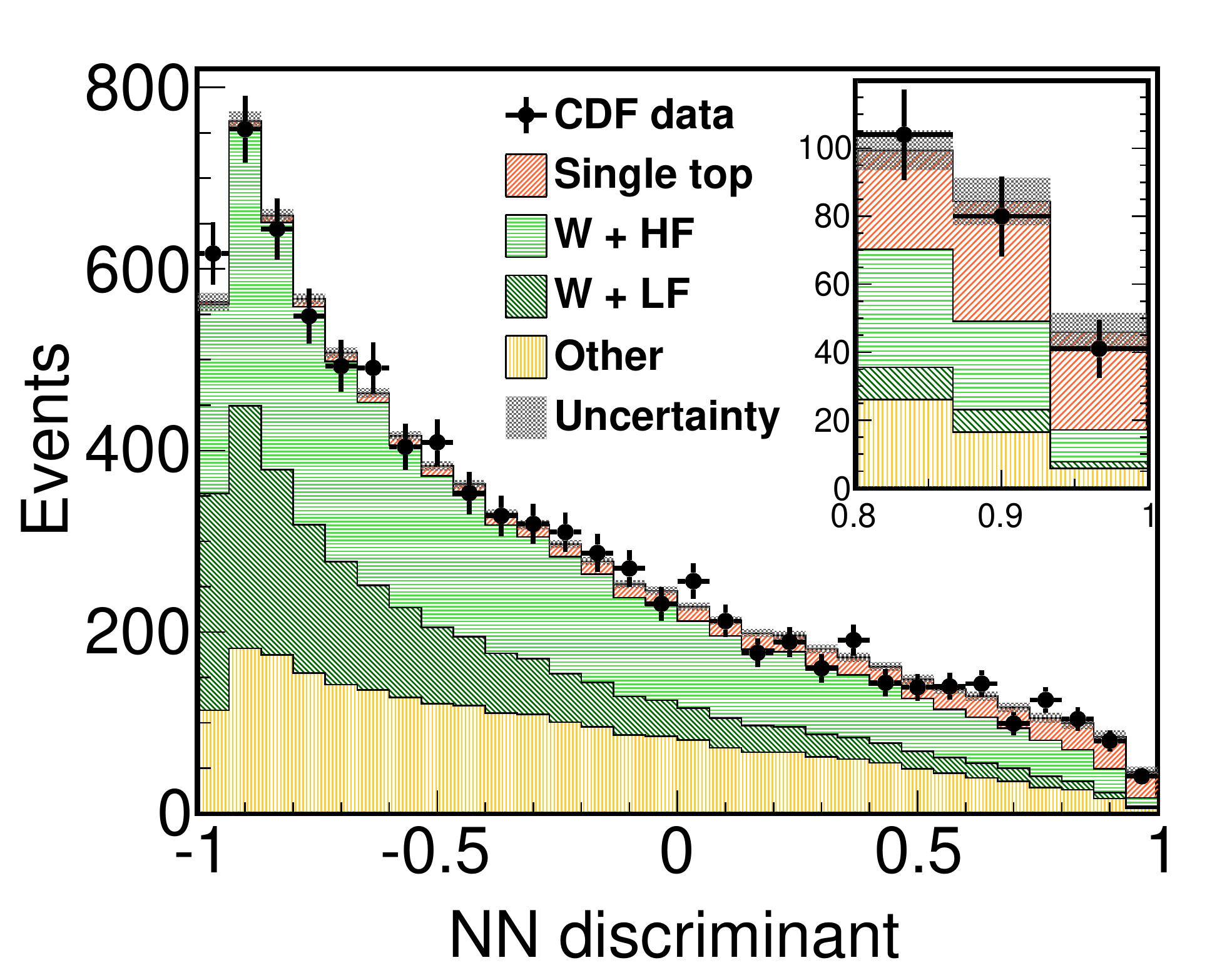}  
\label{fig:all_NNoutSum_inlaidZoom_Stacked_NormPred_PRD}
\caption{\label{fig:all_NN}Comparison of the data with the sum of the
predictions of the NN output for the combined two- and three-jet signal
regions.  The signal + background model is fit to the data.  The uncertainty 
associated with the sum of the predictions (after fitting) is indicated by the grey shaded region 
in each bin.  The inset shows a magnification of the region for which the NN
discriminant ranges from 0.8 to 1.0, where the single top quark contribution
is larger.}
\end{center}  
\end{figure}

The NN output distribution of the combined two- and three-jet signal
regions is shown in Fig.~\ref{fig:all_NN}. The predicted output
distributions of $s$-channel, $t$-channel, and $\Wt$ events are combined
into one signal distribution, with proportions based on the SM predictions.
The measurement of the single top quark cross section is performed using a
maximum posterior density fit to the binned NN output distributions of the
statistically independent bins. We assume a uniform prior probability
density for all non-negative values of the cross section and integrate the
posterior probability density over the parameters of effects associated with
all sources of systematic uncertainties, parametrized using Gaussian
prior-density distributions truncated to avoid negative probabilities.
 
We measure the total cross section of single top quark production
$\sigma_{s+t+\Wt}$, assuming the SM ratio among the $s$-channel,
$t$-channel, and $\Wt$ production rates. From the posterior probability density 
calculated using the NN output distributions, we extract a cross
section of $\sigma_{s+t+\Wt} = 3.04^{+0.57}_{-0.53}$~pb, assuming a top
quark mass of 172.5~GeV/$c^2$.

To extract $\Vtb$, we use the direct proportionality between the production
cross section $\sigma_{s+t+\Wt}$ and $\Vtb^2$~\cite{snote}. We take the
constant of proportionality to be the ratio between the SM prediction for
the cross section $3.40 \pm
0.36$~pb~\cite{PhysRevD.83.091503,PhysRevD.81.054028,PhysRevD.82.054018} and
the nearly unit value of $\Vtb^2$ obtained in the SM assuming the CKM
hierarchy. Under the assumption that the top quark decays to a $W$ boson and
$b$ quark 100\% of the time ($\Vtb^2 \gg \Vts^2 + \Vtd^2$), we obtain a 95\%
Bayesian credibility level lower limit of $\Vtb > 0.78$ and extract $\Vtb =
0.95 \pm 0.09~(\mathrm{stat + syst}) \pm 0.05~(\mathrm{theory})$.

\begin{figure}[tb] 
\begin{center} 
\includegraphics[width=1.0\linewidth, keepaspectratio]{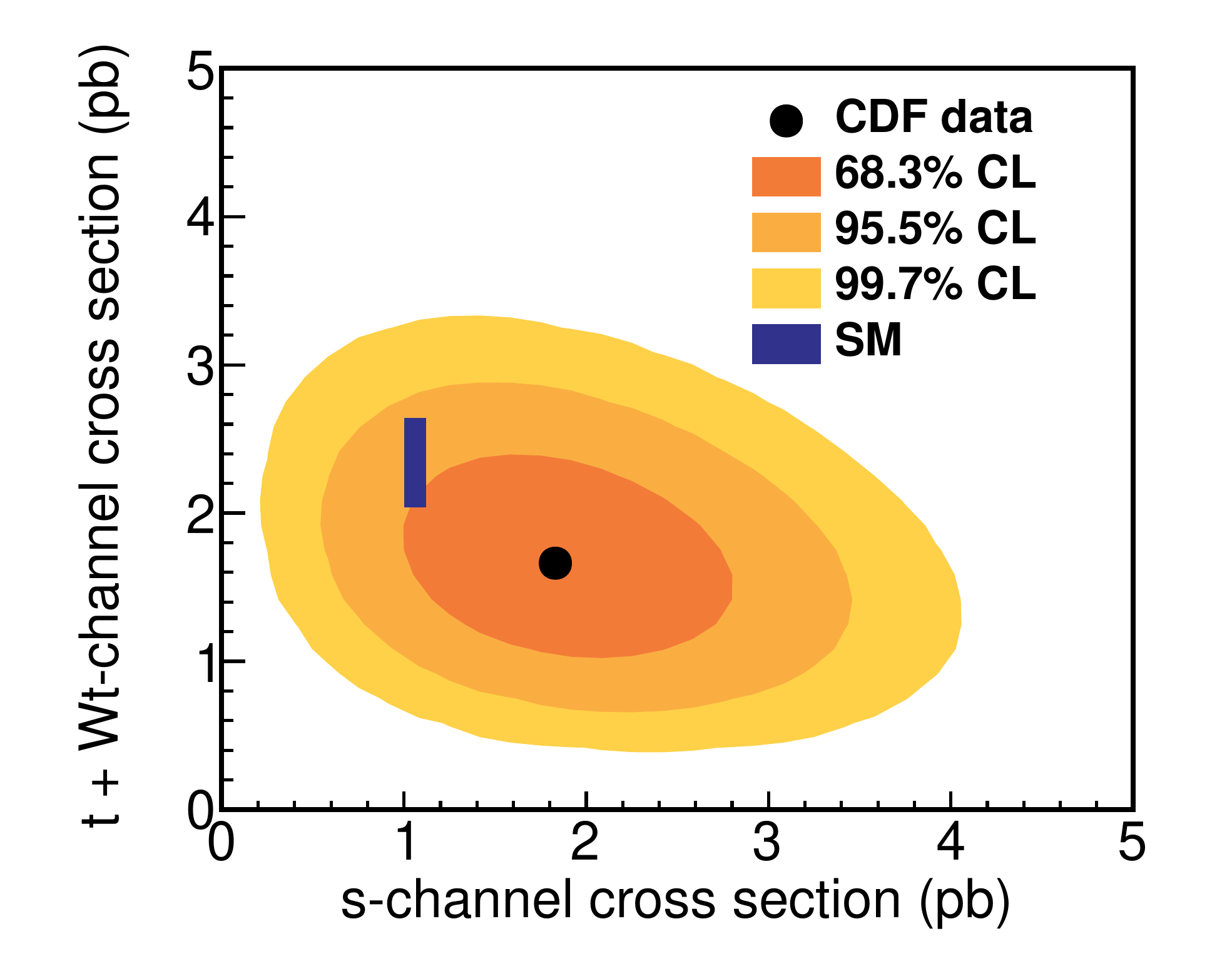} 
\label{fig:2Dplot} 
\caption{\label{fig:2D}Results of the two-dimensional fit for $\sigma_{s}$
and $\sigma_{t+\Wt}$. The black circle shows the best-fit value, and the
68.3\%, 95.5\%, and 99.7\% credibility regions are shown as shaded areas.
The standard model (SM) predictions are also indicated with their
theoretical uncertainties.}
\end{center} 
\end{figure}

To extract the single top quark cross sections for $s$-channel production
and $t$-channel + \Wt production separately, we assume a uniform
prior-probability density distribution in the two-dimensional plane
($\sigma_{s}, \sigma_{t+\Wt}$) and determine the cross sections that
maximize the posterior-probability density distribution. The $t$-channel and
$\Wt$ processes are combined as they share the same final-state topology. We
study the sensitivity of the resulting fit to the relative contribution of
the $t$-channel and $\Wt$ processes (where the $\Wt$ contribution is taken
to be approximately 10\%) and find it to be negligible. The best-fit cross
sections correspond to $\sigma_{s} = 1.81^{+0.63}_{-0.58}$~pb and
$\sigma_{t+\Wt} = 1.66^{+0.53}_{-0.47}$~pb, with a correlation factor of
--24.3\%. The uncertainties on these measurements are correlated because
signal events from both the $s$-channel and the $t$-channel + \Wt processes populate
the signal-like bins of each of our discriminant variables. Regions of
68.3\%, 95.5\%, and 99.7\% credibility are derived by evaluating the
smallest region of area that contains the corresponding fractional integrals
of the posterior-probability density distribution. The best-fit values, the
credibility regions, and the SM predictions are shown in Fig.~\ref{fig:2D}.
These measurements are fully compatible with the SM predictions of
$\sigma_{s}$ = 1.06~$\pm$~0.06~pb and $\sigma_{t+\Wt}$ = 
2.34~$\pm$~0.30~pb~\cite{PhysRevD.83.091503,PhysRevD.81.054028,PhysRevD.82.054018}.
 
In conclusion, we study single top quark production in the $W$ + jets final
state using \ppbar collision data collected by the CDF experiment,
corresponding to 7.5~\invfb of integrated luminosity. We measure a single
top quark cross section for the combined $s$-channel + $t$-channel + $\Wt$
processes of $3.04^{+0.57}_{-0.53}$~pb and we set a lower limit $\Vtb >
0.78$ at the 95\% credibility level, assuming $m_t = 172.5~\gevcc$. Using a
two-dimensional fit for $\sigma_{s}$ and $\sigma_{t+\Wt}$, we obtain
$\sigma_{s} = 1.81^{+0.63}_{-0.58}$~pb and $\sigma_{t+\Wt} =
1.66^{+0.53}_{-0.47}$~pb. All of the measurements are consistent with SM
predictions, and the lower limit on \Vtb places improved bounds on various
extensions of the SM and new phenomena.
 
We thank the Fermilab staff and the technical staffs of the participating
institutions for their vital contributions. This work was supported by the
U.S. Department of Energy and National Science Foundation; the Italian
Istituto Nazionale di Fisica Nucleare; the Ministry of Education, Culture,
Sports, Science and Technology of Japan; the Natural Sciences and
Engineering Research Council of Canada; the National Science Council of the
Republic of China; the Swiss National Science Foundation; the A.P. Sloan
Foundation; the Bundesministerium f\"ur Bildung und Forschung, Germany; the
Korean World Class University Program, the National Research Foundation of
Korea; the Science and Technology Facilities Council and the Royal Society,
United Kingdom; the Russian Foundation for Basic Research; the Ministerio de
Ciencia e Innovaci\'{o}n, and Programa Consolider-Ingenio 2010, Spain; the
Slovak R\&D Agency; the Academy of Finland; the Australian Research Council
(ARC); and the EU community Marie Curie Fellowship Contract No. 302103.


\begin{thebibliography}{43}
\bibitem [{ccm()}]{ccmodes}%
  \BibitemOpen
  \href@noop {} {}\bibinfo {note} {When referring to single top quark
  processes, we implicitly include the charge-conjugate modes.}\BibitemShut
  {Stop}%
\bibitem [{coo()}]{coord}%
  \BibitemOpen
  \href@noop {} {}\bibinfo {note} {CDF uses a cylindrical coordinate system
  where the $z$~axis is along the proton beam direction, $\phi$ is the
  azimuthal angle, and $\theta$ is the polar angle. Pseudorapidity is $\eta =
  -\ln(\tan(\theta/2))$. Transverse momentum is
  $p_{T}=\left|p\right|\sin\theta$ and transverse energy is
  $E_{T}=E\sin\theta$.}\BibitemShut {Stop}%
\bibitem [{cme()}]{cmet2}%
  \BibitemOpen
  \href@noop {} {}\bibinfo {note} {The calorimeter missing transverse energy
  $\METVEC(\mbox{cal})$ is defined as the sum over calorimeter towers
  $\METVEC(\mbox{cal}) = - \sum_{i} E_T^i \hat{n}_i$, where $i$ is an index
  over calorimeter towers with $|\eta| < 3.6$ and $\hat{n}_i$ is a unit vector
  perpendicular to the beam axis and pointing at the $i$th calorimeter tower.
  The reconstructed missing energy $\METVEC$ is derived by correcting
  $\METVEC(\mbox{cal})$ for muon energy deposition and jet energy adjustments.
  We define $\MET(\mbox{cal})$ and $\MET$ to be the scalar magnitudes of
  $\METVEC(\mbox{cal})$ and $\METVEC$, respectively.}\BibitemShut {Stop}%
\bibitem [{dr()}]{dr}%
  \BibitemOpen
  \href@noop {} {}\bibinfo {note} {The distance between the reconstructed
  trajectories of particles or combinations thereof in the $\eta$-$\phi$ plane
  is defined as $R=\sqrt{{\Delta \eta}^2+{\Delta \phi}^2}$.}\BibitemShut
  {Stop}%
\bibitem [{mt()}]{mt}%
  \BibitemOpen
  \href@noop {} {}\bibinfo {note} {The $W$ boson transverse mass is defined as
  $\mtw = \frac{1}{c^{2}}\sqrt{2 E_{T}^{\ell}\slashed
  E_{T}\left(1-\cos\phi_{\ell\nu}\right)}$, where $E_{T}^{\ell}$ is the
  transverse energy of the charged lepton and $\phi_{l\nu}$ is the angle
  between the direction of the charged lepton and the $\slashed
  E_{T}$.}\BibitemShut {Stop}%
\end{thebibliography}
\end{document}